\documentclass[%
reprint, prd
]{revtex4-2}

\usepackage{graphicx}
\usepackage{dcolumn}
\usepackage{bm}
\usepackage{tikz}
\usepackage{amsmath}
\usepackage{amssymb}

\begin{document}

\preprint{APS/123-QED}

\title{The Geometry of Tangent Spaces on Causal Sets}

\author{Samuel Shuman}%
\email{shumans@oregonstate.edu}
\affiliation{Department of Physics, Oregon State University.}

\date{\today}

\begin{abstract}
    In this paper, we expand on previous work describing partial derivatives and metric component estimators to define tangent spaces on causal sets. Partial derivative operators are the basis vectors of the tangent space, and the metric defines the inner product. First, we use partial derivatives of the metric components to define the connection and partial derivatives of the connection to define the curvature. Numerical results show that both of these approach the expected values for a flat spacetime as density increases. Then we used the connection to define parallel transport and geodesics.
\end{abstract}

\maketitle


\section{\label{sec:intro} Introduction}
Casual Set Theory (CST) is a theory of quantum gravity that starts with the assumption that spacetime is discrete. The only structure provided in the theory is a causal order between the events, $\prec$, and the average density of the events in the spacetime, $\rho$. The theory posits that the geometric properties of spacetime can all be recreated from this minimal structure.

There has been significant work estimating geometric properties of causal sets from $\prec$ and $\rho$. For example, the proper time separating two causally connected events can be estimated based on the abundance of chains of different lengths between the events \cite{CausalDiamond}. Similarly, for spacelike separated events, the proper distance can be estimated by looking at the overlap of the events' lightcones \cite{Overlap}. An approach to estimating scalar curvature was discussed in \cite{GeneralizedBox}.

These estimators are tested by using causal sets generated from a background spacetime by a process called sprinkling. To carry out a sprinkling, events are randomly chosen through a Poisson process at a density $\rho$ and the causal order, $\prec$, is inherited from the manifold. You can read more about modeling Poisson processes in \cite{Poisson}.

In General Relativity (GR), spacetime is described in the language of differential geometry. This means that geometric properties are calculated using tangent spaces, which consist of differential operators. In the context of CST, this is difficult to replicate since the theory is discrete. However, recent work estimating partial derivatives and the metric in \cite{myPartialDerivatives} opens the door to a new method for for calculating geometric properties in CST: by representing the tangent spaces directly.

\subsection{Tangent Spaces in General Relativity}
Tangent spaces are a key aspect of differential geometry, and therefore GR. While we cannot go through a detailed introduction to the theory of tangent spaces here, this section should serve as a summary of the important points. 

Like most physical theories, a complete description of GR requires a way to represent vectors, matrices, and other higher-order tensors. In a flat spacetime, vectors can be represented by the separations between events. Since shifting separation vectors from one point to another is trivial in a flat spacetime, it is quite simple to add separations together. In a curved spacetime, however, shifting vectors from one event to another is no longer trivial.

This has two results. First, a curved spacetime will have separate vector spaces defined at each event. These are what we call tangent spaces, since the vectors can be thought of as existing in a flat space that existing tangent to the manifold. Second, the vectors that make up these tangent spaces will not be defined as separation vectors. 

Instead, we use the partial derivative operators along each coordinate as a basis for our vectors. For example, an arbitrary vector $V$ can be written as 
\[V = \sum_\alpha V^\alpha \frac{\partial}{\partial x^\alpha}\]
Here, $x^\alpha$ are the coordinates of our spacetime and $V^\alpha$ are the components of the vector $V$. For convenience, we will define $\partial_\alpha = \frac{\partial}{\partial x^\alpha}$. Also, since many calculations in GR require us to sum over all coordinates, it is common to adopt the Einstein summation notation. This means that any expression which has matching superscripts and subscripts, implies that you are summing over the coordinates. Revisiting our example from earlier, we get
\[V = V^\alpha \partial_\alpha\]
To add or scale such vectors, we simply add or scale the components in the standard way. 

We will also define an inner product between these vectors. The inner product should be a multilinear, symmetric map from pairs of vectors to the real numbers. This is represented by a symmetric matrix with components $g_{\alpha\beta}$ and is called the metric. The inner product is defined by the metric through the equation
\[\langle V, W \rangle = g_{\alpha \beta} V^\alpha W^\beta\]

Now consider the object $g_{\alpha\beta}V^\alpha$. This defines a linear map from vectors to reals. This map is called a dual vector or a covector. The components of the dual vector are defined to be 
\[V_\alpha = g_{\alpha\beta}V^\beta\]
The space of dual vectors has an inner product defined by the inverse metric $g^{\alpha\beta}$, which satisfies $g^{\alpha\beta}g_{\beta\gamma} = \delta^\alpha_\gamma$. Then the inner product can be defined by 
\[\langle V, W \rangle = g_{\alpha\beta}V^\alpha W^\beta = V_\alpha W^\beta = g^{\alpha\beta}V_\alpha W_\beta\]

The process of switching between the vector components and the dual components using the metric is called raising and lowering, due to how it is represented in the indices. The basis for our dual vectors is defined by the inner product 
\[\delta^\alpha_\beta = dx^\alpha \partial_\beta\]

Vectors and covectors are the simplest examples of tensors. In general, a tensor is a multilinear map from vectors and covectors to the reals. A tensor of type $(p,q)$ takes $p$ covectors and $q$ vectors as inputs. For example, a vector is a type $(1,0)$ tensor and a covector is a type $(0,1)$ tensor. The metric is a type $(0,2)$ tensor. 

The components of a tensor can be written using index notation, the same way we have written the components of vectors, covectors, and the metric. For a type $(1,3)$ tensor, such as the Riemann curvature, the components are written as $R^\rho{}_{\sigma\mu\nu}$. The indices on higher-order tensors can be raised and lowered using the metric, just like they were for vectors and covectors
\[R_{\rho\sigma\mu\nu} = g_{\rho\lambda}R^\lambda{}_{\sigma\mu\nu}\]

There are two other operations we need to discuss with tensors, tensor products and tensor contraction. Given two tensors, $A_{\mu\nu}$ and $B^{\alpha\beta}$, the tensor product is found by multiplying components. The result of multiplying a type $(n,m)$ tensor and a type $(p,q)$ tensor is a type $(n+p,m+q)$. 
\[(AB)_{\mu\nu}{}^{\alpha\beta} = A_{\mu\nu}B^{\alpha\beta}\]

Tensor contraction is an operation analogous to taking the trace of the matrix. For example, consider the Riemann curvature tensor $R^\rho{}_{\sigma\mu\nu}$. Contracting this tensor yields the Ricci curvature 
\[R_{\sigma\nu} = R^\lambda{}_{\sigma\lambda\nu} = R^0{}_{\sigma 0 \nu} + R^1{}_{\sigma 1 \nu} + \dots\]
If we then contract the Ricci tensor, we get the scalar curvature.
\[R = R^\nu{}_\nu = g^{\sigma \nu}R_{\sigma \nu}\]

These tensors make up the tangent spaces to our spacetime, and they are an important part of describing the geometry and kinematics of GR. For a manifold $M$, the tangent space to that manifold at a point $x \in M$ is represented symbolically as $T_x M$. The tangent spaces at different events in spacetime are independent, until we define the connection.

\subsubsection{Connection}
In this section we will follow chapter 3 of \cite{Wald} in a discussion of how to define a suitable tensor derivative. This derivative will be denoted $\nabla_\mu$ and is called the covariant derivative. To understand the covariant derivative, we will first list a few properties that we would want such a derivative to have. Let $v^\alpha$ and $w^\alpha$ be vector fields, $T^\alpha{}_\beta$ be a tensor field, and $f$ a scalar function. The covariant derivative should satisfy,

\begin{enumerate}
    \item Linearity: $\nabla_\mu (v^\alpha + w^\alpha) = \nabla_\mu v^\alpha + \nabla_\mu w^\alpha$
    \item Product Rule: $\nabla_\mu (v^\alpha w^\beta) = \nabla_\mu (v^\alpha) w^\beta + v^\alpha \nabla_\mu (w^\beta)$
    \item Commutes with Contractions: $\nabla_\mu T^\alpha{}_\alpha = \delta^\beta_\alpha \nabla_\mu T^\alpha{}_\beta$
    \item Compatible with Partial Derivatives: $\nabla_\mu f = \partial_\mu f$
\end{enumerate}

From these assumed properties, it can be shown that the covariant derivative acts on a vector field as 
\begin{equation}
    \nabla_\mu v^\alpha = \partial_\mu v^\alpha + \Gamma^\alpha{}_{\mu\lambda} v^\lambda \label{covariantDerOnVector}
\end{equation}
For a dual vector, the derivative instead acts as
\begin{equation}
    \nabla_\mu w_\alpha = \partial_\mu w_\alpha - \Gamma^\lambda{}_{\mu\alpha}w_\lambda \label{covariantDerOnDual}
\end{equation}

The $\Gamma$ terms are called the connection or the Christoffel symbols. The first term in equations \ref{covariantDerOnVector} represents taking the derivative of the component functions, while the second term represents taking a derivative of the basis vectors. To apply the covariant derivative to higher order tensors, simply add an appropriate $\Gamma$ term for each index. 
\begin{equation}
    \nabla_\mu T^\alpha{}_\beta = \partial_\mu T^\alpha{}_\beta + \Gamma^\alpha{}_{\mu \lambda}T^\lambda{}_\beta - \Gamma^\lambda{}_{\mu \beta}T^\alpha{}_\lambda \label{covariantDerOnTensor}
\end{equation}

The properties listed above are enough to determine that the covariant derivative must have this type of structure, but we must require two more properties in order to uniquely determine a connection. 
\begin{enumerate}
    \item Torsion Free: $\nabla_\mu \nabla_\nu f = \nabla_\nu \nabla_\mu f$
    \item Metric Compatibility: $\nabla_\mu g_{\alpha\beta} = 0$
\end{enumerate}

To understand the torsion free constraint, consider $\nabla_\alpha \nabla_\beta f$.
\begin{equation}
    \nabla_\alpha \nabla_\beta f = \partial_\alpha \partial_\beta f - \Gamma^\lambda{}_{\alpha\beta} \nabla_\lambda f \label{torsion}
\end{equation}
Since $\nabla_\alpha \nabla_\beta f$ and $\partial_\alpha \partial_\beta f$ are unchanged when $\alpha$ and $\beta$ are swapped, we must have $\Gamma^\lambda{}_{\alpha\beta} = \Gamma^\lambda{}_{\beta\alpha}$. 

Similarly, we can use the metric compatibility condition to further constrain the connection. 
\begin{align}
    &\nabla_\rho g_{\mu\nu} = \partial_\rho g_{\mu\nu} - \Gamma^\lambda{}_{\rho\mu} g_{\lambda\nu} - \Gamma^\lambda{}_{\rho\nu} g_{\mu\lambda} = 0\\
    &\nabla_\mu g_{\nu\rho} = \partial_\mu g_{\nu\rho} - \Gamma^\lambda{}_{\mu\nu} g_{\lambda\rho} - \Gamma^\lambda{}_{\mu\rho} g_{\nu\lambda} = 0\\
    &\nabla_\nu g_{\rho\mu} = \partial_\nu g_{\rho\mu} - \Gamma^\lambda{}_{\nu\rho} g_{\lambda\mu} - \Gamma^\lambda{}_{\nu\mu} g_{\rho\lambda} = 0
\end{align}
Combining these three equations allows you to solve for the connection. 
\begin{equation}
    \Gamma^\alpha{}_{\mu\nu} = \frac{1}{2}g^{\alpha \beta} \big(\partial_\mu g_{\nu \beta} + \partial_\nu g_{\mu\beta} - \partial_\beta g_{\mu\nu}\big) \label{connection}
\end{equation}

This is called the Levi-Cevita connection. Though other connections are sometimes studied in the context of general relativity, the Levi-Cevita connection is the standard choice and will be the connection we estimate on the causal set.

\subsubsection{Curvature \label{sec:R}}
In this section, we will again follow chapter 3 of \cite{Wald} through a brief discussion of the curvature on a manifold. The first representation of curvature we will discuss is the Riemann tensor. There are two alternative ways to define the Riemann tensor which are equivalent on a manifold. One approach is to understand the Riemann curvature as the failure of vectors to return to their original orientation when parallel transported around small loops. We will explore this definition in section \ref{sec:paraR}.

The second way to define the Riemann curvature is by the failure of the covariant derivatives to commute when acting on a covector field. Let $f$ be a scalar field and $w_\mu$ a covector field. Then, by applying the product rule we have
\begin{align}
    &\nabla_\alpha \nabla_\beta (fw_\mu) = (\nabla_\alpha \nabla_\beta f) w_\mu + \nabla_\alpha f \nabla_\beta w_\mu \nonumber\\
    &\hspace{64pt}+ \nabla_\beta f \nabla_\alpha w_\mu + f\nabla_\alpha \nabla_\beta w_\mu \nonumber\\
    &(\nabla_\alpha \nabla_\beta - \nabla_\beta \nabla_\alpha)(fw_\mu) = f(\nabla_\alpha \nabla_\beta - \nabla_\beta \nabla_\alpha)w_\mu \label{DerCommutator}
\end{align}

Since any function can be used for $f$ in equation \ref{DerCommutator}, it must be the case that $(\nabla_\alpha \nabla_\beta - \nabla_\beta \nabla_\alpha)w_\mu$ at a point $p$ depends only on the value of $w_\mu$ at $p$. Thus the action of $(\nabla_\alpha \nabla_\beta - \nabla_\beta \nabla_\alpha)$ can be represented by a $(1,3)$ tensor which we will call the Riemann tensor.
\begin{equation}
    (\nabla_\alpha \nabla_\beta - \nabla_\beta \nabla_\alpha)w_\mu = R^\nu{}_{\alpha\beta\mu}w_\nu \label{RiemannCommutation}
\end{equation}

Using equation \ref{covariantDerOnDual}, we can then calculate the Riemann tensor from the Christoffel symbols.
\begin{equation}
    R^{\rho }{}_{\sigma \mu \nu }=\partial _{\mu }\Gamma ^{\rho }{}_{\nu \sigma }-\partial _{\nu }\Gamma ^{\rho }{}_{\mu \sigma }+\Gamma ^{\rho }{}_{\mu \lambda }\Gamma ^{\lambda }{}_{\nu \sigma }-\Gamma ^{\rho }{}_{\nu \lambda }\Gamma ^{\lambda }{}_{\mu \sigma } \label{RiemannTensor}
\end{equation}
The Riemann tensor has a variety of symmetries and antisymmetries which are discussed in chapter 3 of \cite{Wald}. 

Contracting the Riemann tensor yields another description of the curvature, the Ricci tensor.
\begin{equation}
    R_{\mu\nu} = R^\lambda{}_{\mu \lambda \nu} \label{RicciCurvature}
\end{equation}
Contracting one more time yields the scalar curvature.
\begin{equation}
    R = g^{\mu\nu} R_{\mu\nu} \label{ScalarCurvature}
\end{equation}
The Ricci curvature and scalar curvature are of particular interest because they are included in Einstein's equation relating the curvature of spacetime to the stress-energy tensor $T$. 
\begin{equation}
    R_{\mu\nu} - \frac{1}{2} R g_{\mu\nu} = 8 \pi G T_{\mu\nu}
\end{equation}

\subsection{Functions and Operators}
In order to represent tangent spaces on causal sets, we need representations of scalar functions, such as the coordinates and tensor components, as well as linear operators, such as the partial derivatives. In this paper, we will restrict ourselves to considering functions and operators on finite regions of a causal set. 

A finite region, $A$, of a causal set will have finitely many events in it. Let $n$ be the number of events in $A$. Then a scalar function, $f : A \to \mathbb{R}$, can be specified by $n$ real numbers, that are the values of the function at each event. Therefore, we can represent a function on $A$ as a vector in $\mathbb{R}^n$. 

A linear operator then, should be represented by a linear map acting on $\mathbb{R}^n$. This means such operators can be represented by matrices. 

\subsubsection{Coordinates}
An important set of functions when describing tangent spaces are the coordinates. A set of coordinates on a causal set, denoted $\{x^\alpha\}$, is a set of functions defined in such a way that in the continuum limit, they form a coordinate system on the manifold. 

For a flat spacetime, there has already been work establishing estimators for Cartesian coordinates \cite{JohnstonEmbedding}. In short, the author used proper time estimates defined in \cite{CausalDiamond}, and found the coordinates that best fit these proper times with a Minkowski metric. 

In this paper, we will be focusing on flat spacetimes, but it may be necessary to define coordinates another way if the spacetime is not flat. It is possible, however, that these functions would still define coordinates on a curved spacetime, so long as in the continuum limit, they are still smooth and injective (distinct events must have different coordinates).

\subsubsection{Partial Derivatives and Metric Components}
Partial derivatives are linear operators, so they should be defined by matrices. In \cite{myPartialDerivatives}, two methods for estimating partial derivatives are discussed. Here, we will focus on the more accurate of the two. 

We start with a linearity condition on the derivatives. In the region around some event $e_i$, the change in a function should be explained by the partial derivatives using
\begin{equation}
    \Delta f = \partial_\mu f \Delta x^\alpha
\end{equation}
This condition can be turned into a matrix equation, and the best fit for the partial derivatives can be found using the equation
\begin{equation}
    \begin{bmatrix}
        \partial_0 f\\
        \partial_1 f\\
    \end{bmatrix} = (DX)^+ Df
\end{equation}
Here $D = \mathbb{I} - \delta_i$ and $M^+$ is the Moore-Penrose inverse.

\cite{myPartialDerivatives} carries out a similar process to estimate the metric components. We start with the proper time relation 
\begin{equation}
    \tau^2 = -g_{\alpha\beta}\Delta x^\alpha \Delta x^\beta
\end{equation}
This is then converted to a matrix equation and the Moore-Penrose inverse is used to find the metric components that best fit the proper times estimated as in \cite{CausalDiamond}.

\section{Connection and Curvature}

\subsection{The Levi-Cevita Connection}
The obvious way to define the connection on the causal set is to use equation \ref{connection} with the estimates for the partial derivatives and the metric from \cite{myPartialDerivatives}. It is worth considering, however, that this connection was derived in the continuum by assuming that the covariant derivative must have certain properties. Therefore, we should first test which of these properties can still hold in the context of CST. 

Linearity is an important property of derivatives. Note that even in a causal set equation \ref{covariantDerOnVector} would define a linear derivative operator because the partial derivatives defined in \cite{myPartialDerivatives} were linear. We will therefore require that covariant derivatives still be linear in CST. Similarly, compatibility with partial derivatives is an important enough property that we will impose that as well.

Since the causal set covariant derivative will be compatible with partial derivatives, it cannot follow the product rule. This was shown in section II A of \cite{myPartialDerivatives}, where it was determined that partial derivatives on a causal set cannot always follow the product rule. We also cannot guarantee the covariant derivative will be torsion free, since the partial derivatives are not guaranteed to commute in equation \ref{torsion}.

Thus, even if we define the covariant derivative on a causal set using equation \ref{covariantDerOnVector}, the resulting operator cannot satisfy all of the properties that it has in the continuum. That being said, this definition of the covariant derivative will automatically satisfy metric compatibility and commuting with contractions. Also, since the partial derivatives and the metric defined in \cite{myPartialDerivatives} approach the continuum values as $\rho \to \infty$, properties such as being torsion free and following the product rule should be emergent at high density.

\subsubsection{Numerical Results}
\begin{figure}[h]
    \centering
    \includegraphics[width=8.6cm]{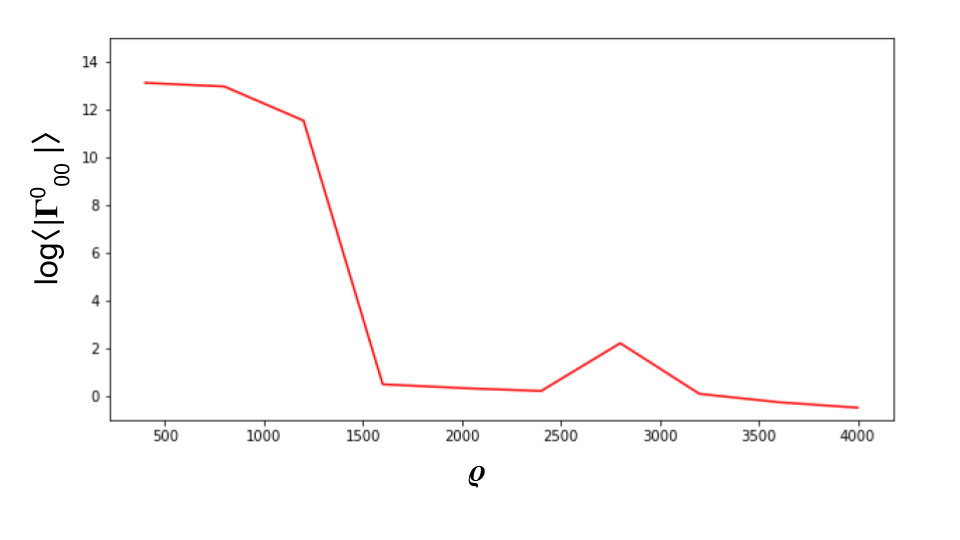}
    \caption{A logarithmic plot of the average value of $|\Gamma^0{}_{00}|$ plotted against the density, $\rho$, of the sprinkling.}
    \label{fig:geoGammaPlot}
\end{figure}

We will carry out numerical tests by sprinkling into a causal diamond in $\mathbb{M}^2$. To test the connection components, note that in the continuum Cartesian coordinates would result in all of the components being zero. In figure \ref{fig:geoGammaPlot}, we see that while $\Gamma^0{}_{00}$ is significantly too large at low densities, it approaches the expected value at higher densities. This suggests that we should expect this approach to give reasonable geometric results, even at densities such as $\rho \sim 4000$. 

Using equations \ref{RiemannTensor}, \ref{RicciCurvature}, and \ref{ScalarCurvature}, it is straightforward to calculate the scalar curvature on a causal set. We should expect the curvature to be 0, since the background spacetime is flat.

\begin{figure*}[ht]
    \centering
    \includegraphics[width=17.2cm]{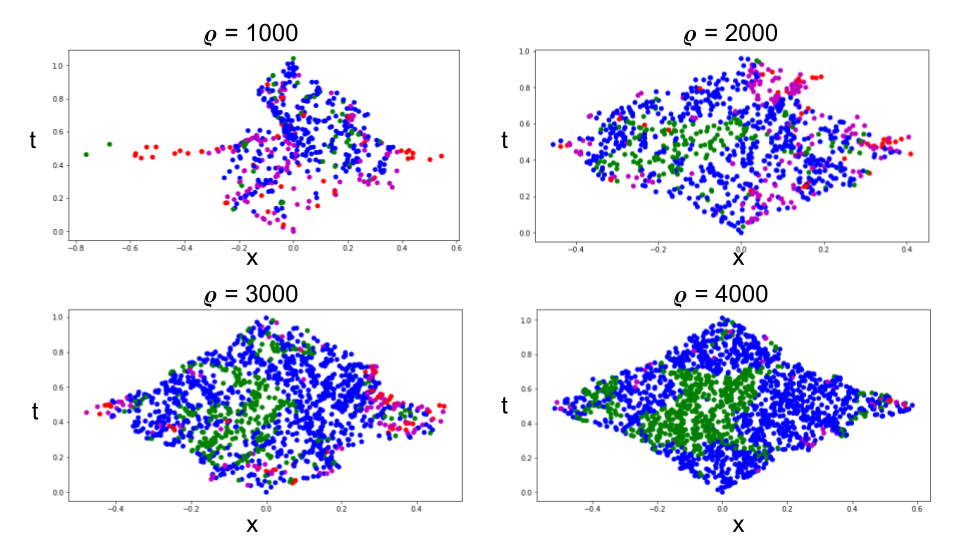}
    \caption{A color plot of the magnitude of the scalar curvature at different densities. Events with $|R| < 10$ are shown in green, those with $|R| < 100$ are shown in blue, those with $|R| < 1000$ are shown in purple, and the rest are shown in red.}
    \label{fig:geoRPlots}
\end{figure*}

In figure \ref{fig:geoRPlots}, we see that the scalar curvature estimated in this way is too large. However, it does seem to approach the expected value as the density increases.

\section{Parallel Transport and Geodesics}

\subsection{Derivatives Along Curves}
The discrete equivalent of a curve is a trajectory, which is a sequence of events in the causal set. First, let us consider how to define tangent vectors to such trajectories. In the continuum, a curve is described by a parameterization of the coordinates on the curve $x^\alpha(\lambda)$. The tangent vectors to the curve are then defined as $v^\alpha = \frac{dx^\alpha}{d\lambda}$. Note that a single curve in the continuum will have many parameterizations. 

To represent a parameterized trajectory in a causal set then, we will not only need a sequence of events, but also a sequence of real numbers $(\lambda_i)$, that corresponds to the value of the parameter $\lambda$ at each event along the trajectory. In particular, the sequence $(\lambda_i)$ must be strictly increasing. To find the tangent vector to a curve in a causal set, we simply approximate the derivative $\frac{dx^\alpha}{d\lambda}$. 
\begin{equation}
    v^\alpha_i = \frac{x^\alpha_{i+1} - x^\alpha_{i}}{\lambda_{i+1} - \lambda_i} = \frac{\Delta x^\alpha}{\Delta\lambda} \label{TangentVector}
\end{equation}

Now we will consider how to take the derivative of scalar function, $f$, along a curve. In the continuum, this can be calculated using the equation
\begin{equation}
    \frac{df}{d\lambda} = \partial_\alpha f \frac{dx^\alpha}{d\lambda} \label{derAlongCurveFromPartials}
\end{equation}
Since we already have representations for partial derivatives and tangent vectors, it is tempting to define derivatives of functions along curves using the above equation. There is a conceptual issue with this representation however. $\frac{df}{d\lambda}$ along a curve should only depend on the value of the function on the curve, while the $\partial_\alpha f$ operators we have defined depend on the value of the function off of the curve as well. Instead, we will directly estimate derivatives of functions along curves the same way we did for tangent vectors.
\begin{equation}
    \frac{df}{d\lambda} \equiv \frac{\Delta f}{\Delta \lambda} \label{derAlongCurve}
\end{equation}

\subsection{Parallel Transport}
Parallel transport is a method for moving a vector along a curve such that the vector does not change. In a flat spacetime, this is as simple as shifting the vector from one event to another, but when the connection is non trivial, a vector's components must change as it moves along a curve in order for the vector to ``not change".

The equation used to rigorously define this condition for a vector $v^\alpha$ is $\frac{D v^\alpha}{d\lambda} \equiv \nabla_\mu v^\alpha \frac{dx^\mu}{d\lambda} = 0$. This can be rewritten in terms of the connection as follows.
\begin{align}
    0 &= \nabla_\mu v^\alpha \frac{dx^\mu}{d\lambda} \nonumber\\
    0 &= (\partial_\mu v^\alpha + \Gamma^\alpha{}_{\mu\nu}v^\nu)\frac{dx^\mu}{d\lambda} \nonumber\\
    \frac{dv^\alpha}{d\lambda} &= -\Gamma^\alpha{}_{\mu\nu}v^\nu\frac{dx^\mu}{d\lambda} \label{PTManifold}
\end{align}

This equation is simple to implement in the causal set, since we have already determined how to represent the connection and derivatives of functions along trajectories. This yields the parallel transport equation for causal set theory, which we will apply in the next two sections.
\begin{equation}
    \Delta v^\alpha = - \Gamma^\alpha{}_{\mu\nu} v^\nu \Delta x^\mu \label{parallelTransport}
\end{equation}

\subsection{Geodesics}
Geodesics are the equivalent of straight lines on a manifold. In general relativity, timelike geodesics describe the paths of objects through spacetime in the absence of other forces, which makes a description of geodesics an important part of the kinematics of any theory of gravity. The characterization of geodesics that we will implement here is that they are trajectories with tangent vectors that are parallel transported along them. 

If we define $v^\alpha_n$ to be the tangent vector to the geodesic at the $n^{\text{th}}$ jump, then we can use equation \ref{parallelTransport} to describe a recursion relation for the tangent vectors. 
\begin{equation}
    v^\alpha_n = v^\alpha_{n-1} - \Gamma^\alpha{}_{\mu\nu} v^\nu_{n-1} \Delta x^\mu \label{geodesicRecurrence}
\end{equation}
Of course, since the causal set is discrete, this tangent vector can only be required to approximately point towards the next event. Thus, we will need to describe a procedure for interpreting this equation in a way that allows us to generate a trajectory in the causal set that will be our geodesic. Recall that a trajectory is a sequence of events, along with a list of values $\lambda$ that corresponds to the parameterization. Then each tangent vector is $v^\alpha = \frac{\Delta x^\alpha}{\Delta \lambda}$. 

To build a geodesic in a causal set, pick an initial event $e_0$, a tangent vector $v^\alpha$, and an initial value of $\lambda$. To find the next event along the trajectory, move along the direction of tangent vector $v^\alpha$ in the coordinate space until you approach ``close enough" to another event that we will call $e_1$. More precisely, $e_1$ is chosen to be the first event to satisfy
\begin{equation}
    ||v^\alpha -  \frac{\Delta x^\alpha}{\Delta \lambda}|| < \varepsilon ||v^\alpha|| \label{geodesicTolerance}
\end{equation}
Here, $\varepsilon$ is the tolerance, which must be chosen, and the next event on the trajectory, $e_1$, is the solution with the smallest positive $\Delta\lambda$. $\lambda(e_1)$ is defined to be $\lambda(e_1) = \lambda(e_0) + \Delta\lambda$. We can then find the new tangent vector using equation \ref{geodesicRecurrence}. This process is repeated to find each event along the curve and the corresponding parameterization. 

The geodesics defined in this way share an important property with their continuum counterparts. Namely, if we scale the initial tangent vector by some factor, the resulting geodesic remains the same. Note that in equation \ref{geodesicRecurrence}, scaling the initial tangent vector by some factor will scale the next tangent vector along the curve by the same factor. Furthermore, in equation \ref{geodesicTolerance}, we will be moving along the same direction in the coordinate space, so the choice for $e_1$ will remain the same as well. The only thing that will change is that the parameterization will be a scaled version of what it was initially.

\subsubsection{Results}

\begin{figure}[h]
    \centering
    \includegraphics[width=8.6cm]{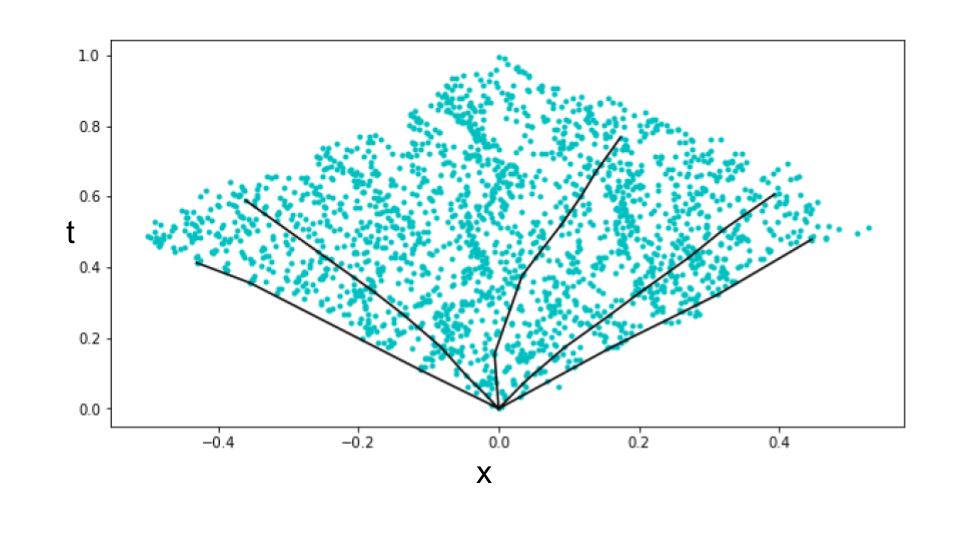}
    \caption{A variety of geodesics on a $\rho = 4000$ sprinkling into $\mathbb{M}^2$. These geodesics were calculated with a tolerance of $\varepsilon = 0.04$.}
    \label{fig:geodesicsPlot}
\end{figure}

In figure \ref{fig:geodesicsPlot}, we see the results of this method for 5 geodesics on a $\rho = 4000$ sprinkling. Since this causal set corresponds to a flat spacetime, we expect these geodesics to be straight lines in the coordinate space. The results do seem to be approximately straight lines, except for the middle geodesic, which curves to the right. This may be happening due to random variations in $\Gamma^\alpha{}_{\mu\nu}$ which result in random variations in $\Delta v^\alpha$ through equation \ref{parallelTransport}. As the density increases, these variations should decrease and the resulting geodesics should approach the continuum geodesics.

\subsection{Curvature From Parallel Transport \label{sec:paraR}}
\begin{figure}[h]
    \centering
    \includegraphics[width=8.6cm]{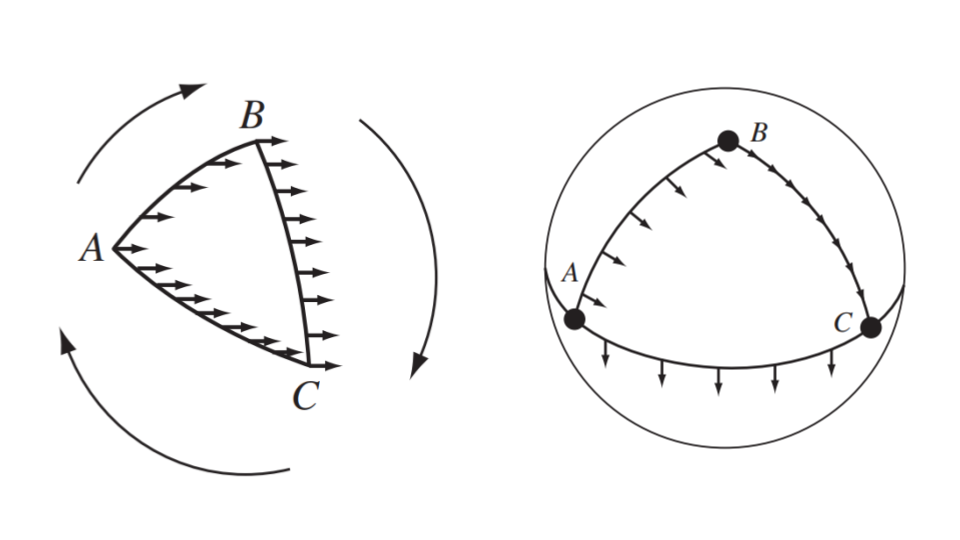}
    \caption{From \cite{Schutz}, a vector is unchanged when parallel-transported around a loop in a flat spacetime. However, in a curved spacetime, a vector is changed when parallel-transported around a loop.}
    \label{fig:PTLoops}
\end{figure}

In section \ref{sec:R}, we discussed a definition of the Riemann curvature in terms of the commutator of covariant derivatives. This definition led to equation \ref{RiemannTensor}.
\[R^{\rho }{}_{\sigma \mu \nu }=\partial _{\mu }\Gamma ^{\rho }{}_{\nu \sigma }-\partial _{\nu }\Gamma ^{\rho }{}_{\mu \sigma }+\Gamma ^{\rho }{}_{\mu \lambda }\Gamma ^{\lambda }{}_{\nu \sigma }-\Gamma ^{\rho }{}_{\nu \lambda }\Gamma ^{\lambda }{}_{\mu \sigma }\]
An equivalent definition is found by considering how vectors are parallel-transported around small loops. In a flat spacetime, a vector is unchanged when parallel-transported around a loop. However, in a curved spacetime, a vector is changed when parallel-transported around a loop (see Figure \ref{fig:PTLoops}). 

In this section we will show that to second-order, a vector parallel-transported around a small triangular loop is changed by
\begin{equation}
    \Delta V^\alpha = \frac{1}{2} A^\mu B^\nu V^\sigma R^\alpha{}_{\mu\nu\sigma} \label{PTLoopR}
\end{equation}
Where $A^\mu$ and $B^\nu$ are the displacement vectors on two sides of the triangle. We will derive this by following the calculation in \cite{Schutz}, which is applied to small rectangular loops. The reason we will derive this relationship for triangular loops is that it is easy to find triangles in a causal set, but the random aspect of sprinklings makes it extremely difficult to find rectangles. Thus, if we want to apply this definition of curvature in a causal set, we must convert the rectangular derivation presented in \cite{Schutz} to a corresponding derivation around a triangular loop. 

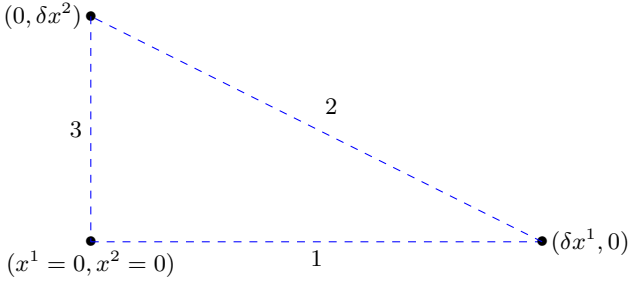
\begin{figure}[h]
    \centering
    \begin{tikzpicture}
        \node[black] at (0,0) {\textbullet};
        \node[below,black] at (0,0) {$(x^1 = 0, x^2 = 0)$};
        \draw [dashed, blue] (0,0) -- (6,0);
        \node[below, black] at (3,0) {$1$};
        \node[black] at (6,0) {\textbullet};
        \node[right,black] at (6,0) {$(\delta x^1, 0)$};
        \draw [dashed, blue] (6,0) -- (0,3);
        \node[right, black] at (3,1.8) {$2$};
        \node[black] at (0,3) {\textbullet};
        \node[left,black] at (0,3) {$(0, \delta x^2)$};
        \draw [dashed, blue] (0,3) -- (0,0);
        \node[left, black] at (0,1.5) {$3$};
    \end{tikzpicture}
    \caption{An arbitrary small triangle on a manifold can be simplified to this setup by choosing a particular coordinate system.}
    \label{fig:triangleSetup}
\end{figure}

To start, consider a small triangular loop. We can always choose a local coordinate system that aligns with the triangle, in the sense that two of the sides will be along the axes of two of the coordinates. This setup is shown in Figure \ref{fig:triangleSetup}.

To calculate the total change in a vector as it is parallel-transported around the triangle, we simply calculate the change in the vector along each side and add it together. 
\begin{equation}
    \Delta V^\alpha = \Delta_1 V^\alpha + \Delta_2 V^\alpha + \Delta_3 V^\alpha
\end{equation}
To calculate how a vector changes along each side, we simply integrate equation \ref{PTManifold}. 
\begin{align}
    \Delta V^\alpha &= \int \frac{dV^\alpha}{d\lambda}d\lambda \nonumber\\
    \Delta V^\alpha &= - \int \Gamma^\alpha{}_{\mu\nu} V^\mu \frac{dx^\nu}{d\lambda} d\lambda
\end{align}

A zeroth-order approximation of the integrand treats $\Gamma^\alpha{}_{\mu\nu}$ and $V^\mu$ as constant around the loop. This results in a change along each side calculated to first-order.
\begin{equation}
    \Delta V^\alpha = - \Gamma^\alpha{}_{\mu\nu} V^\mu \Delta x^\nu \label{firstOrderPT}
\end{equation}
Since $\Delta x^\nu$ is 0 around the loop, the vector does not change to first-order. This is due to the fact that manifolds are locally-flat, which means we will need a second-order approximation to see the effect of curvature. 

To find a second-order approximation of $\Delta V^\alpha$, we will need a first-order approximation of the integrand, $\Gamma^\alpha{}_{\mu\nu} V^\mu$. To first-order, we can approximate the Christoffel symbols with
\begin{align}
    \Gamma^\alpha{}_{\mu\nu}(x^1, x^2) &= \Gamma^\alpha{}_{\mu\nu} + x^1 \partial_1 \Gamma^\alpha{}_{\mu\nu} + x^2 \partial_2 \Gamma^\alpha{}_{\mu\nu} \nonumber\\
    \Gamma^\alpha{}_{\mu\nu}(x^\rho) &= \Gamma^\alpha{}_{\mu\nu} + x^\rho \partial_\rho \Gamma^\alpha{}_{\mu\nu} \label{firstOrderGamma}
\end{align}
Similarly, we can approximate $V^\mu$ to first-order as
\begin{align}
    V^\mu(x^\rho) &= V^\mu + x^\rho \partial_\rho V^\mu \nonumber\\
    V^\mu(x^\rho) &= V^\mu - x^\rho \Gamma^\mu{}_{\sigma\rho}V^\sigma \label{firstOrderV}
\end{align}
The second line in the above equation comes from applying equation \ref{firstOrderPT}. We can now combine these to get a first-order expression for the integrand. After the integral, this results in a second-order expression for $\Delta V^\alpha$. 
\begin{align}
    \Delta V^\alpha = -\int \Gamma^\alpha{}_{\mu\nu}&(V^\nu -\Gamma^\nu{}_{\sigma\rho}V^\sigma x^\rho) \frac{dx^\mu}{d\lambda}d\lambda \nonumber\\
    &-\int x^\rho \partial_\rho \Gamma^\alpha{}_{\mu\nu} V^\nu \frac{dx^\mu}{d\lambda}d\lambda \label{secondOrderPT}
\end{align}

Now we will simply apply equation \ref{secondOrderPT} on each side of the triangle, and add the results to find how the vector changes to second-order around the loop. On side 1, $x^1$ goes from 0 to $\delta x^1$ while $x^2$ is 0. The resulting integral is simple since, in this approximation, $\Gamma^\alpha{}_{\mu\nu}$, $V^\mu$, and $\partial_\rho \Gamma^\alpha{}_{\mu\nu}$ are treated as constants. This results in the following.
\begin{align}
    \Delta_1 V^\alpha = -\Gamma^\alpha{}_{1\nu}V^\nu\delta x^1 &+ \frac{1}{2}\Gamma^\alpha{}_{1\nu}\Gamma^\nu{}_{1\sigma}V^\sigma(\delta x^1)^2 \nonumber\\
    &- \frac{1}{2}\partial_1 \Gamma^\alpha{}_{1\nu} V^\nu(\delta x^1)^2 \label{PTSide1}
\end{align}
Side 3 is also straightforward to calculate, since $x^1$ is 0 while $x^2$ goes from $\delta x^2$ to 0. 
\begin{align}
    \Delta_3 V^\alpha = \Gamma^\alpha{}_{2\nu}V^\nu\delta x^2 &- \frac{1}{2}\Gamma^\alpha{}_{2\nu}\Gamma^\nu{}_{2\sigma}V^\sigma(\delta x^2)^2\nonumber\\
    &+ \frac{1}{2}\partial_2 \Gamma^\alpha{}_{2\nu} V^\nu(\delta x^2)^2 \label{PTSide3}
\end{align}

Side 2 is the most complicated, since both $x^1$ and $x^2$ are changing. We will calculate the change on this side by parameterizing $x^2$ in terms of $x^1$. On this side, $x^2 = \delta x^2 - \frac{\delta x^2}{\delta x^1}x^1$ while $x^1$ goes from $\delta x^1$ to 0. Carrying out the integral yields the following. 
\begin{align}
    \Delta_2 V^\alpha &= \Gamma^\alpha{}_{1\nu}V^\nu \delta x^1 - \Gamma^\alpha{}_{2\nu}V^\nu\delta x^2 - \frac{1}{2}\Gamma^\alpha{}_{1\nu}\Gamma^\nu{}_{1\sigma} V^\sigma (\delta x^1)^2\nonumber\\
    &- \frac{1}{2}\Gamma^\alpha{}_{1\nu}\Gamma^\nu{}_{2\sigma} V^\sigma \delta x^1 \delta x^2 + \frac{1}{2}\Gamma^\alpha{}_{2\nu}\Gamma^\nu{}_{1\sigma} V^\sigma \delta x^1 \delta x^2 \nonumber\\
    &+ \frac{1}{2}\Gamma^\alpha{}_{2\nu}\Gamma^\nu{}_{2\sigma} V^\sigma (\delta x^2)^2 + \frac{1}{2} \partial_1 \Gamma^\alpha{}_{1\nu}V^\nu(\delta x^1)^2 \nonumber\\
    &- \frac{1}{2} \partial_2 \Gamma^\alpha{}_{2\nu}V^\nu(\delta x^2)^2 + \frac{1}{2} \partial_2 \Gamma^\alpha{}_{1\nu} V^\nu \delta x^1 \delta x^2\nonumber\\
    &- \frac{1}{2} \partial_1 \Gamma^\alpha{}_{2\nu} V^\nu \delta x^1 \delta x^2 \label{PTSide2}
\end{align}
Adding together the change in the vector for each side yields,
\begin{align}
    \Delta V^\alpha &= \Delta_1 V^\alpha + \Delta_2 V^\alpha + \Delta_3 V^\alpha \nonumber\\
    \Delta V^\alpha &= \frac{1}{2}\delta x^1 \delta x^2 V^\sigma (\partial_2 \Gamma^\alpha{}_{1\sigma} - \partial_1 \Gamma^\alpha{}_{2\sigma}\nonumber\\
    &\hspace{64pt}+ \Gamma^\alpha{}_{2\rho}\Gamma^\rho{}_{1\sigma} - \Gamma^\alpha{}_{1\rho}\Gamma^\rho{}_{2\sigma}) \nonumber\\
    \Delta V^\alpha &= \frac{1}{2} A^\mu B^\nu V^\sigma R^\alpha{}_{\mu\nu\sigma} \label{PTTotal}
\end{align}

\subsubsection{Implementation on a Causal Set}
The preceding derivation demonstrates how the change in a vector around a small loop is related to the curvature. Though we have previously defined the Riemann curvature directly in terms of partial derivatives of the Christoffel symbols, in this section, we will explore the possibility of defining the curvature in terms of parallel transport around loops. 

To setup this calculation, pick an event $e \in \mathcal{C}$, and consider all other pairs of events in a small coordinate ball around $e$. If we start with a basis vector $V^\alpha = \delta^\alpha_i$, we can calculate how it changes moving around the triangular loop using parallel transport. We can also represent $A^\mu$ and $B^\nu$ as the displacements from $e$ to the other events in the loop. Using this, we now have all the pieces to represent equation \ref{PTTotal} as a matrix equation, which will allow us to find the best-fit for the curvature using the Moore-Penrose inverse. In two-dimensions, this matrix equation looks like the following. 

\begin{align}
    &\begin{bmatrix}
        (\Delta V^0)_1 & (\Delta V^1)_1\\
        (\Delta V^0)_2 & (\Delta V^1)_2\\
        \vdots & \vdots\\
        (\Delta V^0)_k & (\Delta V^1)_k
    \end{bmatrix}
    = \nonumber\\
    &\frac{1}{2}\begin{bmatrix}
        (A^0B^0)_1 & (A^1B^0 - A^0B^1)_1 & (A^1B^1)_1\\
        (A^0B^0)_2 & (A^1B^0 - A^0B^1)_2 & (A^1B^1)_2\\
        \vdots & \vdots & \vdots\\
        (A^0B^0)_k & (A^1B^0 - A^0B^1)_k & (A^1B^1)_k
    \end{bmatrix}
    \begin{bmatrix}
        R^0{}_{00i} & R^1{}_{00i}\\
        R^0{}_{01i} & R^1{}_{01i}\\
        R^0{}_{11i} & R^1{}_{11i}
    \end{bmatrix}
\end{align}
Here, $k$ is the number of distinct pairs of events near $e$ and $i$ is the basis vector chosen. Also note that this equation has been designed to enforce that $R^\alpha{}_{\mu\nu\sigma} = - R^\alpha{}_{\nu\mu\sigma}$, in the same way that equation 28 in \cite{myPartialDerivatives} enforced that the metric was symmetric. If we call the matrix on the left $\Delta$, and the matrices on the right $(AB)$ and $R$ respectively, then we can solve for $R$ using the Moore-Penrose inverse, which will find the least-squares solution to the equation.
\begin{equation}
    R = 2(AB)^+\Delta \label{paraREqn}
\end{equation}
Note that this does not solve for all of the components of $R$, instead it solves for all $R^\alpha{}_{\mu\nu i}$, where $i$ is the chosen basis vector. This process would have to be repeated with each basis vector to find the full Riemann tensor. 

\begin{figure*}[ht]
    \centering
    \includegraphics[width = 17.2cm]{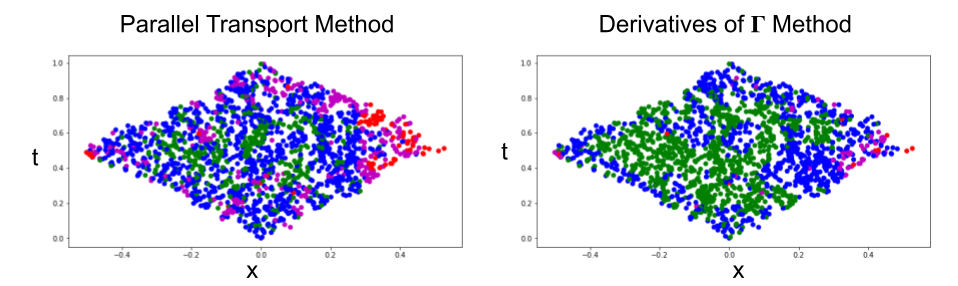}
    \caption{A color plot of the magnitude of the scalar curvature $|R|$ calculated at a density of $\rho = 4000$ using two methods. The plot on the left shows the results using equation \ref{paraREqn} with coordinate balls of size $\varepsilon = 0.05$ while the plot on the right show the results using equation \ref{RiemannTensor}. Events with $|R| < 10$ are shown in green, those with $|R| < 100$ are shown in blue, those with $|R| < 1000$ are shown in purple, and the rest are shown in red.}
    \label{fig:paraRPlot}
\end{figure*}

In figure \ref{fig:paraRPlot}, we see the results for the scalar curvature using equation \ref{paraREqn}, compared to the results using equation \ref{RiemannTensor}. While there are many events with $|R| < 100$ in the plot on the left, suggesting that calculating $R$ from parallel transport may be feasible, it is also clear that the results using equation \ref{RiemannTensor} are significantly better.

\section{\label{sec:conc}Conclusions}
This paper built on previous work estimating partial derivatives and metric components to formulate a representation of tangent spaces on causal sets. The resulting connection, curvature, and geodesics were tested numerically for a causal diamond in $\mathbb{M}^2$, and showed promising results. 

The main limitation of this approach is that tests were only done to densities up to $\rho = 4000$, and only in a flat spacetime. An obvious direction for future work is to continue these tests at higher densities and in curved spacetimes. 

Another interesting direction for future research is to develop a theory of perturbations on causal sets. Now that we have a description of tangent spaces, we have the necessary tools to describe perturbations to the metric, which are solutions to the linearized Einstein equations. This approach could also provide a preliminary description of gravitational waves on causal sets by describing wavelike perturbations. 
\newpage

\bibliography{tangentGeo}

\end{document}